\documentclass[10pt]{iopart}

\usepackage{graphicx}
\usepackage{siunitx}
\usepackage{caption, subcaption}
\usepackage{xcolor}
\usepackage{bm}

\usepackage{xcolor}

\usepackage[urlcolor=blue,colorlinks=true,citecolor=blue,linkcolor=blue,pdfstartview={FitH},bookmarks=false]{hyperref}

\usepackage{cite}
\begin{document}

\title[]{Evidences for local non-centrosymmetricity and strong phonon anomaly in EuCu$_{2}$As$_{2}$:\\A Raman spectroscopy and lattice dynamics study}
\author{Debasmita Swain$^{1}$, Mainak Dey Sarkar$^{2}$, Andrzej Ptok$^{3}$, G.~Vaitheeswaran$^{2}$, Anushree Roy$^{1}$, Sitikantha D Das$^{1}$}

\address{$^1$Department of Physics, Indian Institute of Technology Kharagpur, Kharagpur 721302, India}
\address{$^2$School of Physics, University of Hyderabad, Gachibowli, Hyderabad 500046, Telangana, India}
\address{$^3$ Institute of Nuclear Physics, Polish Academy of Sciences, W. E. Radzikowskiego 152, PL-31342 Krak\'{o}w, Poland}
\ead{aptok@mmj.pl}
\ead{vaithee@uohyd.ac.in}
\ead{anushree@phy.iitkgp.ac.in}
\ead{sitikantha.das@iitkgp.ac.in}

\vspace{10pt}

\date{\today}

\begin{abstract}
Phonon modes and their association with the electronic states have been investigated for the metallic EuCu$_{2}$As$_{2}$ system. 
In this work, we present the Raman spectra of this pnictide system which clearly shows the presence of seven well defined peaks above $100$~cm$^{-1}$ that is consistent with the locally non-centrosymmetric {\it P4/nmm} crystal structure, contrary to that what is expected from the accepted symmorphic {\it I4/mmm} structure. 
Lattice dynamics calculations using the {\it P4/nmm} symmetry attest that there is a commendable agreement between the calculated phonon spectra at the $\Gamma$ point and the observed Raman mode frequencies, with the most intense peak at $\sim 232$~cm$^{-1}$ being ascribed to the A$_{1g}$ mode. 
Temperature dependent Raman measurements show that there is a significant deviation from the expected anharmonic behaviour around $165$~K for the A$_{1g}$ mode, with anomalies being observed for several other modes as well, although to a lesser extent. 
Attempts are made to rationalize the observed anomalous behavior related to the hardening of the phonon modes, with parallels being drawn from metal dichalcogenide and allied systems.
Similarities in the evolution of the Raman peak frequencies with temperature seem to suggest a strong signature of a subtle electronic density wave instability below $165$~K in this compound.
\end{abstract}

\maketitle

\section{Introduction}

The issue as to whether crystallographic inversion symmetry is present or is broken locally has important consequences, as it can lead to the emergence of novel physical phenomena. 
The majority of 122 compounds, including the first heavy-fermion superconductor CeCu$_{2}$Si$_{2}$~\cite{steglihc.aarts.79, Stockert2011} or iron-based superconductors $A$Fe$_{2}$As$_{2}$ (where $A$ is alkali metal)~\cite{PhysRevB.78.014523, Li_2020}, have a tetragonal ThCr$_{2}$Si$_{2}$-like structure ({\it I4/mmm} which respects inversion symmetry).
However, there are several examples of 122 systems, which are realized in CaBe$_{2}$Ge$_{2}$-like structures ({\it P4/nmm} symmetry, where inversion symmetry is broken locally).
Such crystal structure is rather unusual in the context of 122 family.
Nevertheless, we can mention several examples, like SrPt$_{2}$As$_{2}$~\cite{kudo.nishikubo.10}, $R$Pt$_{2}$Si$_{2}$ (where $R=$Y, La, Nd, and Lu)~\cite{nagano.araoka.13}, LaPd$_{2}$Bi$_{2}$~\cite{mu.pan.18}, or recently discovered CeRh$_{2}$As$_{2}$~\cite{kim.landaeta.21}.
These systems show a broad range of interesting phenomena, viz. superconductivity, magnetic order, or density waves~\cite{kim.landaeta.21,rotter.tegel.08,sasmal.lorenz.08,torikachvili.budko.08,zocco.grube.13,cho.yang.17}.

\begin{figure}[!b]
\centering
\includegraphics[width=0.7\linewidth]{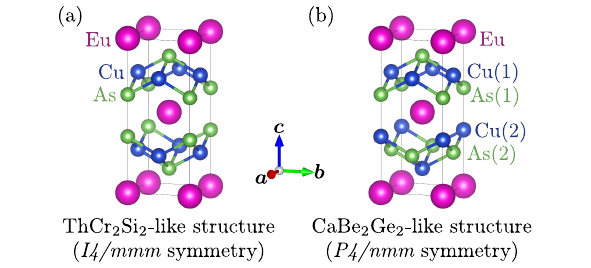}
\caption{
Crystal structure of EuCu$_{2}$As$_{2}$ in two polymorphous phases: (a) ThCr$_{2}$Si$_{2}$-like structure ({\it I4/mmm} symmetry), and (b) CaBe$_{2}$Ge$_{2}$-like structure ({\it P4/nmm} symmetry).
} 
\label{fig.crys}
\end{figure}

Although both the phases of 122 systems have an overall inversion symmetry in the crystal structure, the position of the center of symmetry is not the same. The origin of this is the difference in the stacking order of the block layers. (Fig.~\ref{fig.crys}).
For example, EuCu$_{2}$As$_{2}$ with {\it I4/mmm} possess two blocks with the same order of atoms (i.e., As--Cu--As).
Contrary to this, in the case of {\it P4/nmm} symmetry, Eu layers are located between two different blocks (Cu--As--Cu and As--Cu--As).
In some situations, the type of structure ({\it I4/mmm} or {\it P4/nmm}) can depend on the valence electron counts~\cite{lin.miller.14}. Furthermore, since there are two Cu ions which occupy different crystallographic locations, their relative contributions to the electronic states are expected to be distinct.
Also, some chemical similarities between compounds can suggest what structure can be more realistic~\cite{wang.botti.21}.
Nevertheless, the question about ``true'' crystal structure in some situations is still open.

In the context of the Cu-based 122 compounds of the form $A$Cu$_2$As$_2$ ($A=$Sr, Ba, Eu), it is known that they generically belong to the {\it I4/mmm} phase, with enhanced covalency due to interlayer As--As bonding. 
This results in a ``collapsed'' tetragonal phase, leading to a $sp$ band metallic phase with relatively weak electronic correlations~\cite{singh.09}. 
Recent ARPES measurements~\cite{wu.richard.15} of BaCu$_2$As$_2$ have, however, shown the presence of surface states, possibly arising from the structural reconstruction associated with the strain relaxation at the surface of the bulk collapsed phase. 
Additionally, Raman measurements have pointed out the presence of unidentified extra peaks in the spectra, inconsistent with the bulk {\it I4/mmm} symmetry~\cite{wu.richard.15}.

The puzzle as to whether a particular 122 compound will form in the {\it I4/mmm} or {\it P4/nmm} structure is significant crystallographically and also has important consequences on the type of electronic or magnetic phases that can be stabilized. 
For example, the electronic phase diagram of the Fe-based 122 systems crystallising mostly in the {\it I4/mmm} structure, comprises of spin density waves, unconventional superconductivity, and nematicity originating predominantly from the electron dynamics within the iron pnictide layers~\cite{Das.Craco.15}. 
On the other hand, other types of 122 pnictides (e.g., CeRh$_{2}$As$_{2}$) crystallising in the locally non-centrosymmetric {\it P4/nmm} structure can host charge~\cite{ptok.kapcia.21} and quadrupole~\cite{Hafner.Khanenko.22} density wave phases arising primarily from the transition metal-pnictide interlayer interaction.

Previous studies on EuCu$_{2}$As$_{2}$ have mainly focused on the nature of magnetic ordering of the Eu sublattice, around $17.5$~K~\cite{sengupta.paulose.05,sangupta.alzamora.12,anand.johnston.15}. 
However, in the context of iron-base 122 superconductors (e.g., EuFe$_{2}$As$_{2}$), there has been a significant debate about whether the electron dynamics of the transition metal--pnictide layer are at all affected by the localized $f$ states of Eu . 
While some reports suggest that the Eu ordering temperature is significantly robust against the pressure and chemical dilution in the FeAs layer~\cite{miclea.nicklas.09}, thereby suggesting negligible coupling between the Eu and FeAs sublattice, other photoemission studies confirm significant hybridization between the localized Eu $f$ states with the As $p$ orbitals~\cite{Adhikary_2013}. 
Nevertheless, the strongly localized $f$ states of Eu can have an effect on the physical properties even above the ordering temperature, i.e., in the paramagnetic phase.
Thus, the role of Eu magnetic moment on influencing the dynamics of the electronic bands is still a matter of debate.

In order to address these issues, we perform  micro Raman spectroscopic measurements on EuCu$_{2}$As$_{2}$ and complement the experimental findings with lattice dynamic calculations. Furthermore, temperature dependent studies indicate the presence of non trivial electronic dynamics in the system. In the absence of any prior reports of such an investigation on this system, we find significant evidence for {\it P4/nmm} structure. This is in stark contrast to the established {\it I4/mmm} structure that is believed to be the correct crystal phase for all Cu-As based 122 systems.

The paper is organized as follows:
Used techniques are described in Sec.~\ref{sec.tech}.
Next, we present our experimental and theoretical results in Sec.~\ref{sec.res}.
In particular, we present the system characterization in Sec.~\ref{sec.system}, discuss the Raman spectrum and lattice dynamics in Sec.~\ref{sec.raman}. The temperature dependence of the Raman spectral profile and  phonon anomaly are addressed in Sec.~\ref{sec.temp}.
Finally, we summarize the paper with the main conclusions in Sec.~\ref{sec.summary}.

\section{Techniques}
\label{sec.tech}

{\it Experimental details.}
Powder X-ray diffraction (XRD) patterns of EuCu$_2$As$_2$ were analyzed using the Fullprof software following the Rietveld profile refining method, and crystal structure was drawn using VESTA Software. 
Energy dispersive X-ray analysis (EDAX) also confirms the measured stoichiometry of Eu, Cu and As very close to the expected value (see Fig.~SF1 of the Supplementary Material (SM)).
DC magnetic susceptibility measurements were carried out in the temperature range of $10$--$300$~K in the presence of applied magnetic field of $H=100$~Oe employing a Quantum Design (QD) superconducting quantum interference device (SQUID). 
Micro-Raman spectroscopy in the back-scattering geometry was carried out using LabRam HR Evolution (Horiba France). 
Spectra were recorded using $532$~nm laser excitation wavelength. 
For this measurement, we utilized a 50X objective with a spot size of 2$\mu$$m$. 
In order to avoid the heating effect of the laser power on the sample, the laser power on the sample was kept $< 1$~mW. 
Temperature dependent Raman measurements were carried out using a sample stage (THMS-600, Linkam, UK) over the temperature range between $100$~K and $300$~K.

{\it Computational details.}
The first-principles density functional theory (DFT) calculations were performed using the projector augmented-wave (PAW) potentials~\cite{blochl.94} implemented in 
the Vienna Ab initio Simulation Package ({\sc Vasp}) code~\cite{kresse.hafner.94}.
For the exchange-correlation energy, the generalized gradient approximation (GGA) in the Perdew, Burke, and Ernzerhof for solids (PBEsol) parametrization was used~\cite{perdew.ruzsinszky.08}.
The strong correlation effects on the Eu $f$ state treated as a valence state were introduced within DFT+U scheme, proposed by Dudarev {\it et al.}~\cite{dudarev.botton.98}, with $U = 7$~eV.
The energy cutoff for the plane-wave expansion was set to $400$~eV.
The optimization of the lattice constant and atomic positions in the presence of the spin--orbit coupling were performed for conventional unit cell (containing two formula units), using $12 \times 12 \times 6$ {\bf k}--point grid, using the Monkhorst--Pack scheme~\cite{monkhorst.pack.76}.
As a convergence condition of the optimization loop, we took the energy change below $10^{-6}$~eV and $10^{-8}$~eV for ionic and electronic degrees of freedom.
The symmetry of the system was analyzed by {\sc FindSym}~\cite{stokes.hatch.05} and {\sc Spglib}~\cite{togo.tanaka.18}, while momentum space analyses were performed within {\sc SeeK-path}~\cite{hinuma.pizzi.17}.

The dynamical properties were calculated using the supercell technique, with (extended) direct {\it Parlinski--Li--Kawazoe} method~\cite{parlinski.li.97}.
In this calculation, the inter-atomic force constants (IFC) were found from the force acting on each of the atoms.
Here, each atom was displaced in random directions.
In calculations, we used the 20 samples of supercell containing $2\sqrt{2} \times 2\sqrt{2} \times 1$ conventional cells (i.e., supercell is rotated $45^{\circ}$ around $c$ axis with respect to the conventional cell and contains 80 atoms).
The forces acting on each atom were calculated via {\sc Vasp}, and reduce the $4 \times 4 \times 4$ ${\bm k}$-grid.
The IFC was calculated using {\sc alamode} software~\cite{tadano.gohda.14}.
Analysis of the irreducible representations of phonon modes at $\Gamma$ point were performed using {\sc phonopy} package~\cite{togo.chaput.23}.

\begin{figure}[!b]
\centering
\includegraphics[width=0.8\linewidth]{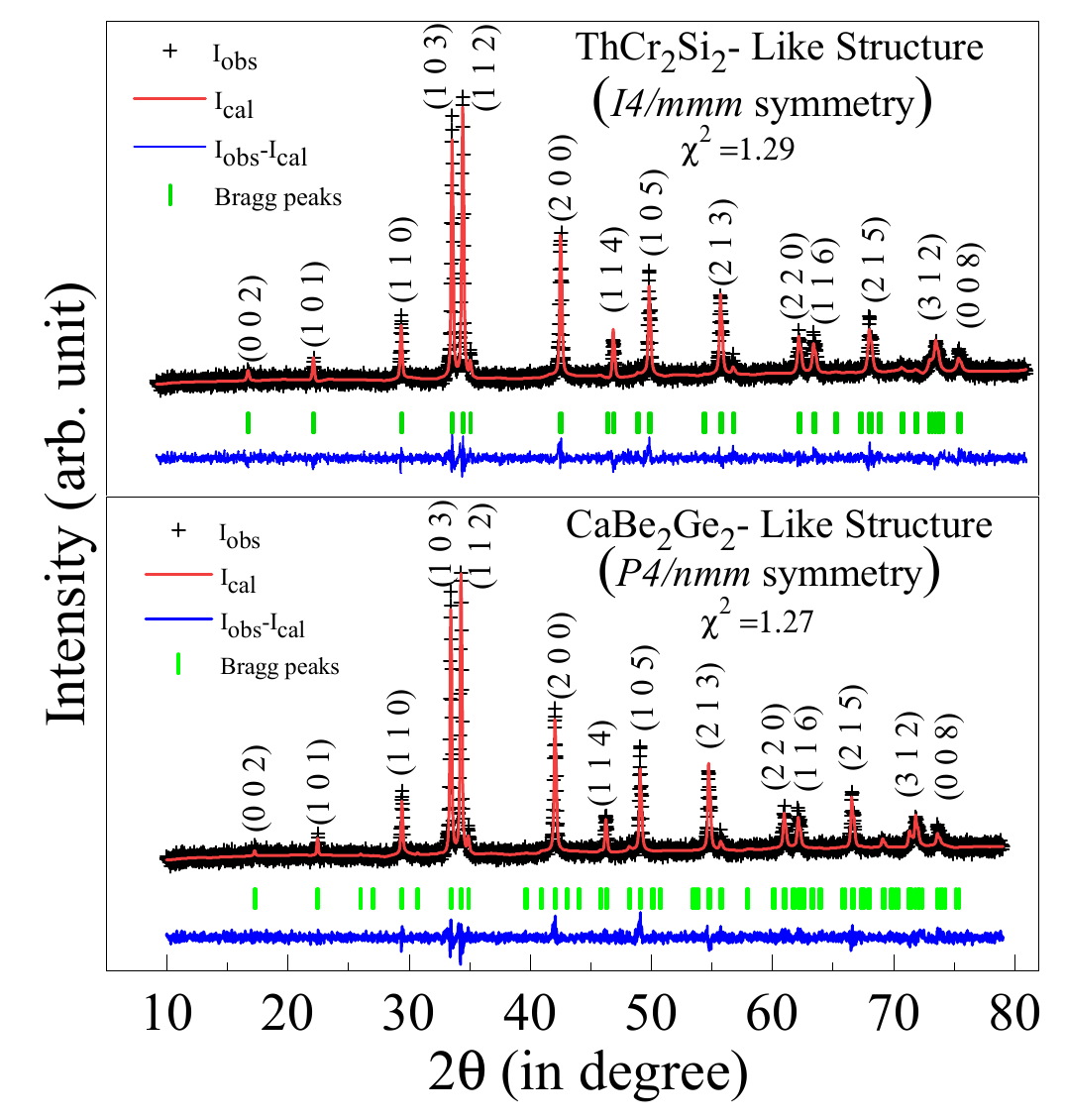}
\caption{Powder X-ray diffraction pattern (with Cu - K$_\alpha$ radiation, $\lambda$ = 1.5418 \si{\angstrom}) of the EuCu$_2$As$_2$ sample (black crosses) and its Rietveld refinement (red curve) in the case of {\it I4/mmm} (top) and {\it P4/nmm} (bottom) symmetry.} 
\label{fig.xrd}
\end{figure}
\section{Results and discussions}
\label{sec.res}

\subsection{System characterization}
\label{sec.system}

As we mentioned before, EuCu$_{2}$As$_{2}$ was initially recognized as {\it I4/mmm} symmetry~\cite{sengupta.paulose.05,sangupta.alzamora.12,anand.johnston.15}. Powder XRD data of our sample shows the single phase nature of the compound without any signature of impurity phases.
The obtained refined XRD pattern of the As-grown sample following the Rietveld profile method can be reproduced by both {\it I4/mmm} and {\it P4/nmm} symmetry (see Fig.~\ref{fig.xrd}). The fact that the earlier accepted space group being {\it I4/mmm} is well borne out by the fact that the sum of the $(hkl)$ values are even for all the observed intensity peaks, whereas an odd value would signal the presence of {\it P4/nmm}.
It can be seen that all the diffraction peaks can be very well indexed with both symmetries. Thus, the obtained XRD pattern cannot be used as an indisputable argument for confirming any of these symmetries, especially if the odd valued peaks have a very low intensity profile, although however, a very slow scan specifically to look for them might indicate their existence and thereby reveal the correct symmetry.

\begin{table}[!t]
\centering
\scriptsize
\caption{
\label{tab.crystal} 
{Comparison of the obtained (experimental and theoretical) lattice parameters of EuCu$_{2}$As$_{2}$ with previously reported results. XRD data was collected at 300 K.}}

\begin{tabular}{ccccc}
\hline
\hline
\multicolumn{5}{c}{EuCu$_{2}$As$_{2}$} \\
\hline
$a$ & \multicolumn{3}{c}{$4.215(1)$~\AA} & Ref.~\cite{dunner.mewis.95} \\
 & \multicolumn{3}{c}{$4.2330(1)$~\AA} & Ref.~\cite{anand.johnston.15} \\
 & \multicolumn{3}{c}{$4.29(7)$~\AA} & Ref.~\cite{sengupta.paulose.05} \\
 & \multicolumn{3}{c}{$4.29(7)$~\AA} & {\bf This work} (Exp) \\
(for P4/nmm) & \multicolumn{3}{c}{$4.235$~\AA} & {\bf This work} (DFT) \\
(for I4/mmm) & \multicolumn{3}{c}{$4.163$~\AA} & {\bf This work} (DFT) \\
\hline
$c$ & \multicolumn{3}{c}{$10.185(2)$~\AA} & Ref.~\cite{dunner.mewis.95} \\
 & \multicolumn{3}{c}{$10.1683(3)$~\AA} & Ref.~\cite{anand.johnston.15} \\
 & \multicolumn{3}{c}{$10.185(1)$~\AA} & Ref.~\cite{sengupta.paulose.05} \\
 & \multicolumn{3}{c}{$10.28(8)$~\AA} & {\bf This work} (Exp) \\
(for P4/nmm) & \multicolumn{3}{c}{$9.920$~\AA} & {\bf This work} (DFT) \\
(for I4/mmm) & \multicolumn{3}{c}{$10.988$~\AA} & {\bf This work} (DFT) \\
\hline
\multicolumn{5}{c}{{\bf P4/nmm}} \\
\hline
Atom & Eu & $2c$ & (1/4,1/4,0.2460) & {\bf This work} (DFT) \\
positions & Cu(1) & $2a$ & (3/4,1/4,0) & \\
& Cu(2) & $2c$ & (1/4,1/4,0.6277) & \\
& As(1) & $2c$ & (1/4,1/4,0.8688) & \\
& As(2) & $2b$ & (3/4,1/4,1/2) & \\
\hline
Atom & Eu & $2c$ & (1/4,1/4,0.2512) & {\bf This work} (Exp) \\
positions & Cu(1) & $2a$ & (3/4,1/4,0) & \\
& Cu(2) & $2c$ & (1/4,1/4,0.6227) & \\
& As(1) & $2c$ & (1/4,1/4,0.8665) & \\
& As(2) & $2b$ & (3/4,1/4,1/2) & \\
\hline
\multicolumn{5}{c}{{\bf I4/mmm}} \\
\hline
Atom & Eu & $2a$ & (0,0,0) & {\bf This work} (DFT) \\
positions & Cu & $4d$ & (0,1/2,1/4) & \\
& As & $4e$ & (0,0,0.3753) & \\
\hline
Atom & Eu & $2a$ & (0,0,0) & {\bf This work} (Exp) \\
positions & Cu & $4d$ & (0,1/2,1/4) & \\
& As & $4e$ & (0,0,0.3787) & \\
\hline
\hline
\end{tabular}
\end{table}

The Rietveld refinement on the XRD pattern yields lattice constants, which are compared with previously reported in Tab.~\ref{tab.crystal}.
The lattice parameters obtained theoretically for EuCu$_{2}$As$_{2}$ reproduce the experimental ones quite well, while atomic positions are similar to that reported in other {\it P4/nmm} 122 compounds (e.g.,~EuPd$_{2}$Sb$_{2}$~\cite{das.mcfadden.10}).

\begin{figure}[!t]
\centering
\includegraphics[scale=0.55]{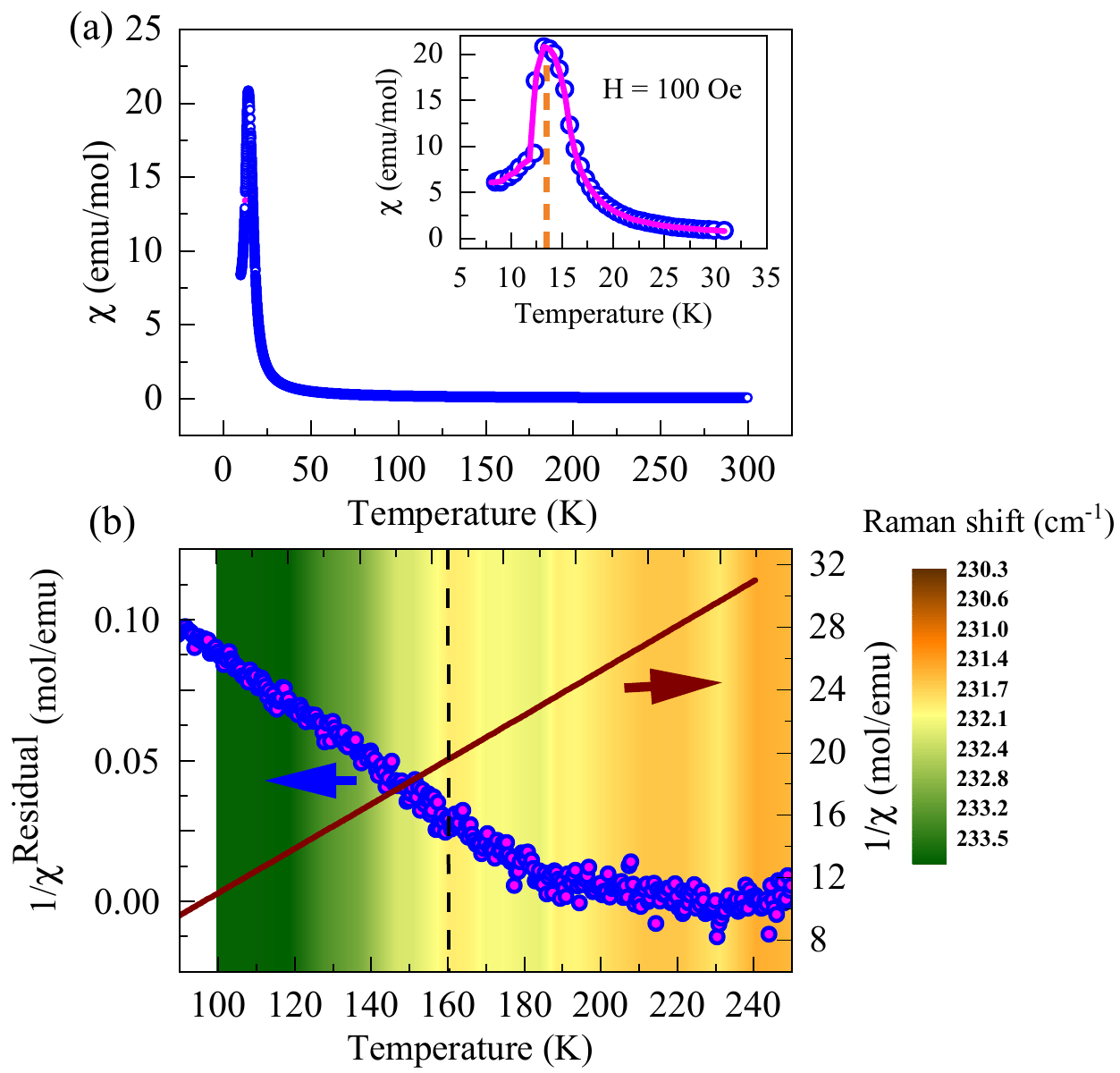}
\caption{(a) Magnetic susceptibility plot of EuCu$_2$As$_2$ within a temperature range from $10$ to $300$~K, showing Eu$^{2+}$ ions magnetic ordering around $14$~K (see inset). (b) Temperature dependence of inverse magnetic susceptibility of EuCu$_2$As$_2$ (black solid line) with the Curie--Weiss fitting (red solid curve), (right axis) Residual plot after subtraction of the Curie--Weiss contribution of Eu$^{2+}$ magnetic moments (left axis). Colors in the plot represent evolution of the Raman shift with temperature of EuCu$_2$As$_2$.}
 \label{fig.raman1s}
\end{figure}

Additionally, magnetic measurements confirmed the presence of Eu ordering around $\sim 14$~K~\cite{sengupta.paulose.05,sangupta.alzamora.12,anand.johnston.15}, as shown in the magnetisation versus temperature plot Fig.~\ref{fig.raman1s}(a). 
To probe any contribution of spin degrees of freedom in EuCu$_2$As$_2$ for governing  the observed  anomaly of the A$_{1g}$ phonon mode, we analysed the high temperature susceptibility of the system. 
The black line in Figure~\ref{fig.raman1s}(b) shows the inverse magnetic susceptibility ($1/{\chi}$) plot of EuCu$_2$As$_2$ over the temperature range between $100$~K and $250$~K. 
Needless to mention that at first glance, it appears that the plot follows the Curie--Weiss law, $1/\chi=(T+\theta)/C$, where $C$ is the Curie--Weiss constant and $\theta$ is the Curie temperature. 
However, in view of the observed anomaly in the phonon mode of the system below $165$~K in Fig.~7(b) of the paper, we fitted the data points using the above Curie--Weiss relation above $180$~K, and extrapolated down to $100$~K.
The determined effective magnetic moment is $7.74$~$\mu_B$ per formula unit, closely approaching the anticipated theoretical value of $7.94$~$\mu_B$ for a free Eu$^{2+}$ ion. 
After subtraction of the obtained Curie--Weiss contribution from the data, an upturn in ${1/\chi}$ can be observed below $165$~K, which is shown in Figure~\ref{fig.raman1s}(b). 
It is to be noted that if the Curie--Weiss law prevailed over the entire range of temperature, one would have expected only a fluctuation in data points below $165$~K within the noise level. 
Similar deviation from the standard Curie--Weiss fit has been observed in related 122 systems~\cite{rotter.tegel.08b,krellner.caroca.08}, where it has been ascribed to SDW transition. 
The contour plot of the Raman shift with temperature, as displayed in Fig.~\ref{fig.raman1s}(b), vividly portrays the evolution observed in the magnetization plot.

\subsection{Structural symmetry assignments}
\label{sec.raman}

\begin{figure}[!b]
\centering
\includegraphics[width=0.8\linewidth]{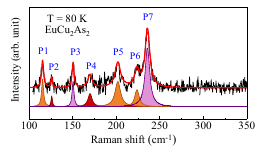}
\caption{
Background subtracted Raman spectra of EuCu$_2$As$_2$ at $80$~K. 
All  peaks are fitted with  Lorentzian profile (cumulative fit is given by solid red lines and individual mode fit by the corresponding filled colour lines). Peaks of same colour correspond to a particular symmetry assignment -- magenta: A$_{1g}$, orange: E$_{g}$, red: B$_{1g}$.} 
\label{fig.low_temp}
\end{figure}

\begin{figure*}
\centering
\includegraphics[width=1.1\linewidth]{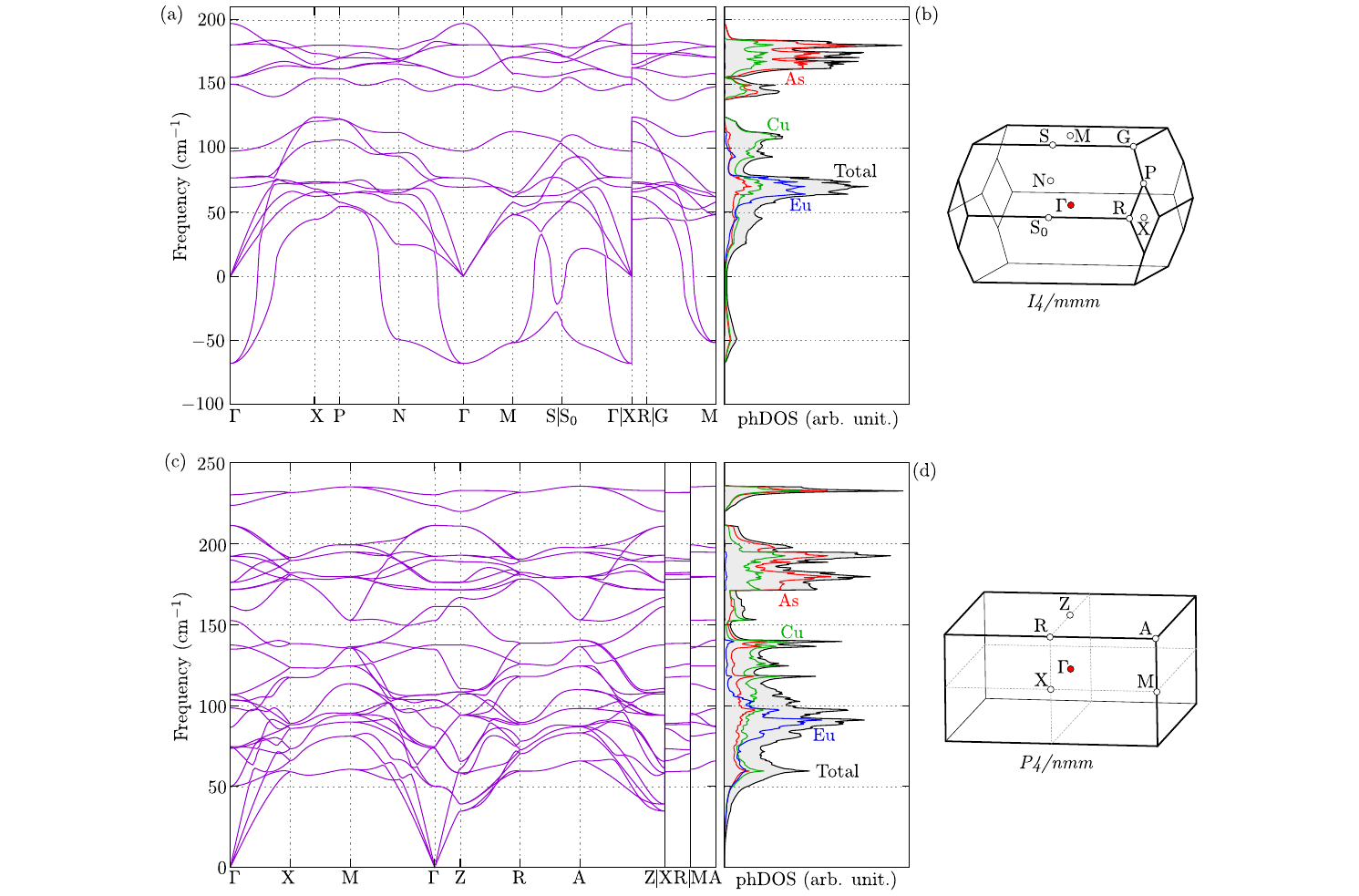}
\caption{Phonon dispersion curves (a,c) and phonon density of states (b,d) for EuCu$_{2}$As$_{2}$ with {\it I4/mmm} and {\it P4/nmm} (top and bottom panels, respectively).
Presented results were obtained in the presence of AFM order.
Right panels present the Brillouin zone and their high symmetry points for both symmetries.}
\label{fig.ph_band}
\end{figure*}

Fig.~\ref{fig.low_temp} plots the linear background subtracted Raman spectra of EuCu$_2$As$_2$, recorded at $80$~K.
The spectrum was deconvoluted with a sum of Lorentzian functions $\sum_{i} A_i / [ ( \omega - \omega_0 )^{2} + \Gamma^2 ]$, where $A_i$, $\omega_i$, and $\Gamma_i$ are integral intensity, peak position, and width of the $i$th mode in the spectrum.
There are seven visible peaks at 
 $115$~cm$^{-1}$ (P1), $126$~cm$^{-1}$ (P2), $150$~cm$^{-1}$ (P3), $170$~cm$^{-1}$ (P4), $201$~cm$^{-1}$ (P5), $224$~cm$^{-1}$ (P6), and $235$~cm$^{-1}$ (P7) (see Fig.~\ref{fig.low_temp}).

The experimental results can be compared with the lattice dynamics study obtained theoretically.
We calculated the phonon dispersion curves and phonon density of states (DOS) for EuCu$_{2}$As$_{2}$ with {\it I4/mmm} and {\it P4/nmm} symmetry [top and bottom panels in Fig.~\ref{fig.ph_band}, respectively].
As we can see, in the case of {\it I4/mmm} symmetry the system is dynamically unstable due to the appearance of imaginary soft modes [presented as negative frequencies in Fig.~\ref{fig.ph_band}(a)].
Such imaginary soft modes are associated with the vibrations of Cu and As in $ab$ plane.
Contrary to this, for {\it P4/nmm}, the phonon spectrum does not exhibit any imaginary soft mode [Fig.~\ref{fig.ph_band}(b)]. 
Hence it can be concluded that the system is stable with {\it P4/nmm} symmetry.
Interestingly, the maximal range of frequencies are much smaller for {\it I4/mmm} symmetry (around $200$~cm$^{-1}$)  than that for {\it P4/nmm} symmetry.
It is to be noted that the observed Raman peak around $230$~cm$^{-1}$ cannot be explained by calculations for system with {\it I4/mmm} symmetry. We should also notice that the primitive unit cell of {\it P4/nmm} is twice bigger than this related for {\it I4/mmm}. This is reflected in the phonon dispersion curve by increasing the number of phonon modes [cf.~Fig.~\ref{fig.ph_band}(a) and~\ref{fig.ph_band}(c)], as well as a larger number of the available Raman active modes.

Finally, it is observed that the phonon DOS for both symmetries exhibit similar features [cf.~Fig.~\ref{fig.ph_band}(b) and~\ref{fig.ph_band}(d)].
It is interesting to note that the vibrations of Eu atoms dominate at lowest frequencies (below $\sim 150$~cm$^{-1}$), mostly around $75$~cm$^{-1}$.
The phonon modes at highest frequencies (above $\sim 150$~cm$^{-1}$) are primarily associated with Cu and As atom vibrations.
In this range of frequencies, vibrations of Eu are mostly not realized.

\begin{table}[!t]
\centering
\scriptsize
\caption{
\label{tab.freq} 
Irreducible representations (Irrep.) and characteristic frequencies (cm$^{-1}$) of the phonon modes at the $\Gamma$ point for EuCu$_{2}$As$_{2}$ with {\it P4/nmm} symmetry. 
Bolded values correspond to the Raman active modes.
Results in the case of non-magnetic (NM) and antiferromagnetic (AFM) states.
}
\begin{tabular}{cccc}
\hline
\hline
\multicolumn{4}{c}{EuCu$_{2}$As$_{2}$} \\
\multicolumn{4}{c}{({\it P4/nmm})} \\
Irrep. & Theo. (NM) & Theo. (AFM) & Exp (80 K) \\
\hline
E$_\text{g}$ & \textbf{44.00} & \textbf{50.10} & N/A \\
E$_\text{g}$ & \textbf{70.48} & \textbf{73.68} & N/A \\
E$_\text{u}$ & 71.32 & 74.59 & \\
A$_\text{1g}$ & \textbf{76.32} & \textbf{86.49} & N/A \\
A$_\text{2u}$ & 89.16 & 98.54 & \\
E$_\text{u}$ & 95.43 & 103.47 & \\
E$_\text{g}$ & \textbf{100.27} & \textbf{106.74} & 115 (P1) \\
A$_\text{2u}$ & 129.69 & 134.83 & \\
B$_\text{1g}$ & \textbf{130.42} & \textbf{137.40} & 126 (P2)\\
A$_\text{1g}$ & \textbf{151.91} & \textbf{152.51} & 150 (P3) \\
B$_\text{1g}$ & \textbf{159.31} & \textbf{161.21} & 170 (P4) \\
E$_\text{u}$ & 170.02 & 171.65 & \\
E$_\text{u}$ & 174.02 & 176.06 & \\
A$_\text{2u}$ & 187.80 & 189.96 & \\
E$_\text{g}$ & \textbf{190.26} & \textbf{192.23} & 201 (P5) \\
E$_\text{g}$ & \textbf{210.31} & \textbf{211.28} & 224 (P6)\\
A$_\text{2u}$ & 223.39 & 223.65 & \\
A$_\text{1g}$ & \textbf{230.93} & \textbf{230.23} & 235 (P7) \\
\hline
\end{tabular}
\end{table}

\begin{table}[!t]
\centering
\scriptsize
\caption{
\label{tab.freq2} 
Irreducible representations (Irrep.) and characteristic frequencies (cm$^{-1}$) of the phonon modes at the $\Gamma$ point for EuCu$_{2}$As$_{2}$ with {\it I4/mmm} symmetry.
Results for AFM phase.
Bolded values correspond to the Raman active modes.
Negative frequency correspond to the imaginary soft mode (see also Fig.~\ref{fig.ph_band}(a)).
}
\begin{tabular}{cc|cc}
\hline
\hline
\multicolumn{4}{c}{EuCu$_{2}$As$_{2}$} \\
\multicolumn{4}{c}{({\it I4/mmm} -- unstable)} \\
$\quad$ Irrep. & Freq. & $\quad$ Irrep. & Freq. \\
\hline
E$_\text{g}$ & \textbf{-50.07i} & A$_\text{2u}$ & 163.01 \\
A$_\text{2u}$ & 86.26 & E$_\text{u}$ & 173.62 \\
E$_\text{u}$ & 92.20 & E$_\text{g}$ & \textbf{192.17} \\
B$_\text{1g}$ & \textbf{117.15} & A$_\text{1g}$ & \textbf{199.94} \\
\hline
\end{tabular}
\end{table}

{\it Irreducible representations at $\Gamma$ point.}
Now, we present the theoretical analysis of the phonon modes at $\Gamma$ point.
In the case of {\it I4/mmm} symmetry, the phonon modes at $\Gamma$ point can be decomposed into irreducible representations as follows:
\begin{eqnarray*}
\label{eq.irr1}
\Gamma_\text{acoustic} &=& \text{A}_\text{2u} + \text{E}_\text{u} \\
\nonumber \Gamma_\text{optic} &=& \text{A}_\text{1g} + 2 \text{A}_\text{2u} + \text{B}_\text{1g} + 2 \text{E}_\text{u} + 2 \text{E}_\text{g} .
\end{eqnarray*}
Here, A$_\text{2u}$ and E$_\text{u}$ are infra-red 
(IR) active, while modes A$_\text{1g}$, B$_\text{1g}$ and E$_\text{g}$ are Raman active. 
Similarly, for {\it P4/nmm} symmetry, we get:
\begin{eqnarray*}
\label{eq.irr2}
\Gamma_\text{acoustic} &=& \text{A}_\text{2u} + \text{E}_\text{u} \\
\nonumber \Gamma_\text{optic} &=& 3 \text{A}_\text{1g} + 4 \text{A}_\text{2u} + 2 \text{B}_\text{1g} + 4 \text{E}_\text{u} + 5 \text{E}_\text{g}
\end{eqnarray*}
where A$_\text{2u}$ and E$_\text{u}$ are IR active, while A$_\text{1g}$, B$_\text{1g}$, and E$_\text{g}$ are Raman active.
As we can see, the number of Raman active modes is more in the case of {\it P4/nmm} system with respect to the {\it I4/mmm} structure.
This is a consequence of the difference in the number of atoms in primitive unit cells for both structures, mentioned earlier.
Nevertheless, in both cases, A$_\text{1g}$, A$_\text{2u}$, and B$_\text{1g}$ are associated with vibrations of the atoms along the $c$ direction.
Similarly, double-degenerate modes E$_\text{u}$ and E$_\text{g}$ are related to the vibrations of  atoms in the $ab$ plane.

Irreducible representations and characteristic frequencies of the phonon modes at $\Gamma$ point are tabulated in Tab.~\ref{tab.freq} and Tab.~\ref{tab.freq2}, for {\it P4/nmm} and {\it I4/mmm} symmetries, respectively.
We also compare frequencies of Raman active modes with those observed experimentally (see Fig.~\ref{fig.low_temp}).
By comparing peaks from P1 to P7, we can recognize them as Raman active modes with symmetries E$_\text{g}$, B$_\text{1g}$, A$_\text{1g}$, B$_\text{1g}$, E$_\text{g}$, E$_\text{g}$, and A$_\text{1g}$ modes (for {\it P4/nmm} symmetry), respectively.
As we can see, theoretically obtained phonon frequencies at $\Gamma$ point for EuCu$_{2}$As$_{2}$ quite well reproduce the experimentally found frequencies, confirming the realization of {\it P4/nmm} structure by this compound (Tab.~\ref{tab.freq}). 
Contrary to this, the results obtained for {\it I4/mmm} exhibit earlier mentioned soft mode and what is more are mismatch with respect to the experimental data (Tab.~\ref{tab.freq2}).

It is worth reminding ourselves that Raman is essentially a near surface probe. Thus, the presence of Raman peaks greater than four in number, expected for {\it I4/mmm} symmetry, would experimentally negate the presence of this phase at least on the surface. Purely from experiments it is possible to envisage a situation where the surface belongs to the {\it P4/nmm} phase and the bulk belongs to the {\it I4/mmm} phase. However, from DFT calculations that were performed in this case were essentially for the bulk so that it would confirm that the whole structure belongs to the {\it P4/nmm} phase.

\begin{figure}[!t]
\centering
\includegraphics[width=0.8\linewidth]{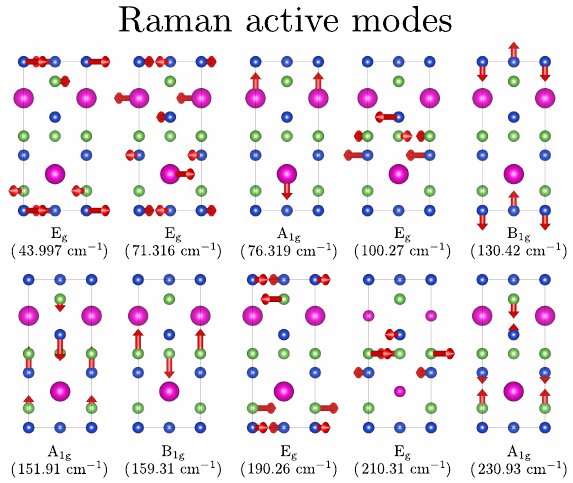}
\caption{Atoms displacement realized by the Raman active modes for EuCu$_2$As$_2$ with {\it P4/nmm} symmetry.}
\label{fig.active}
\end{figure}

Visualization of the atoms displacement realized by the Raman active modes for EuCu$_{2}$As$_{2}$ with {\it P4/nmm} symmetry are presented in Fig.~\ref{fig.active} (the atoms displacement by IR active modes are available in Fig.~SF2 in the SM).
In the case of {\it I4/mmm} symmetry (not presented; examples of the atoms displacement in the case of KFe$_{2}$As$_{2}$ with {\it I4/mmm} are available in Ref.~\cite{litvinchuk.hadjiev.08,wdowik.jaglo.15,baum.li.18,ptok.sternik.19}), the A$_\text{1g}$ mode can be realized only by As atoms, the B$_\text{1g}$ only by Cu atoms, while E$_\text{g}$ modes by both Cu and As atoms.
Here, Eu atoms do not participate in the Raman active modes.
Contrary to this, atomic displacement patterns realized by Raman active modes for systems with {\it P4/nmm} symmetry is much ``richer''.
A$_\text{1g}$ modes can be realized only by atoms in the Wyckoff position $2c$ [i.e., Eu, Cu(2), and As(1)], B$_\text{1g}$ modes can be realized by only atoms in the Wyckoff position $2a$ and $2b$ [i.e., Cu(1), and As(2)], while E$_\text{g}$ modes by all type of atoms (cf.~atom positions are marked in Fig.~\ref{fig.crys}). The mentioned difference arise from the two types of Cu--As block in {\it I4/mmm} and {\it P4/nmm} structures (cf. structures presented in Fig.~\ref{fig.crys}).
In the case of {\it I4/mmm} the mirror symmetry in $ab$ plane does not change the crystal structure, and Raman active modes are the same in both Cu--As blocks.
Such operation is not permitted in the case of {\it P4/nmm} structure, where two types of Cu--As--Cu and As--Cu--As are realized.

\begin{figure*}
\centering
\includegraphics[width=0.89\linewidth]{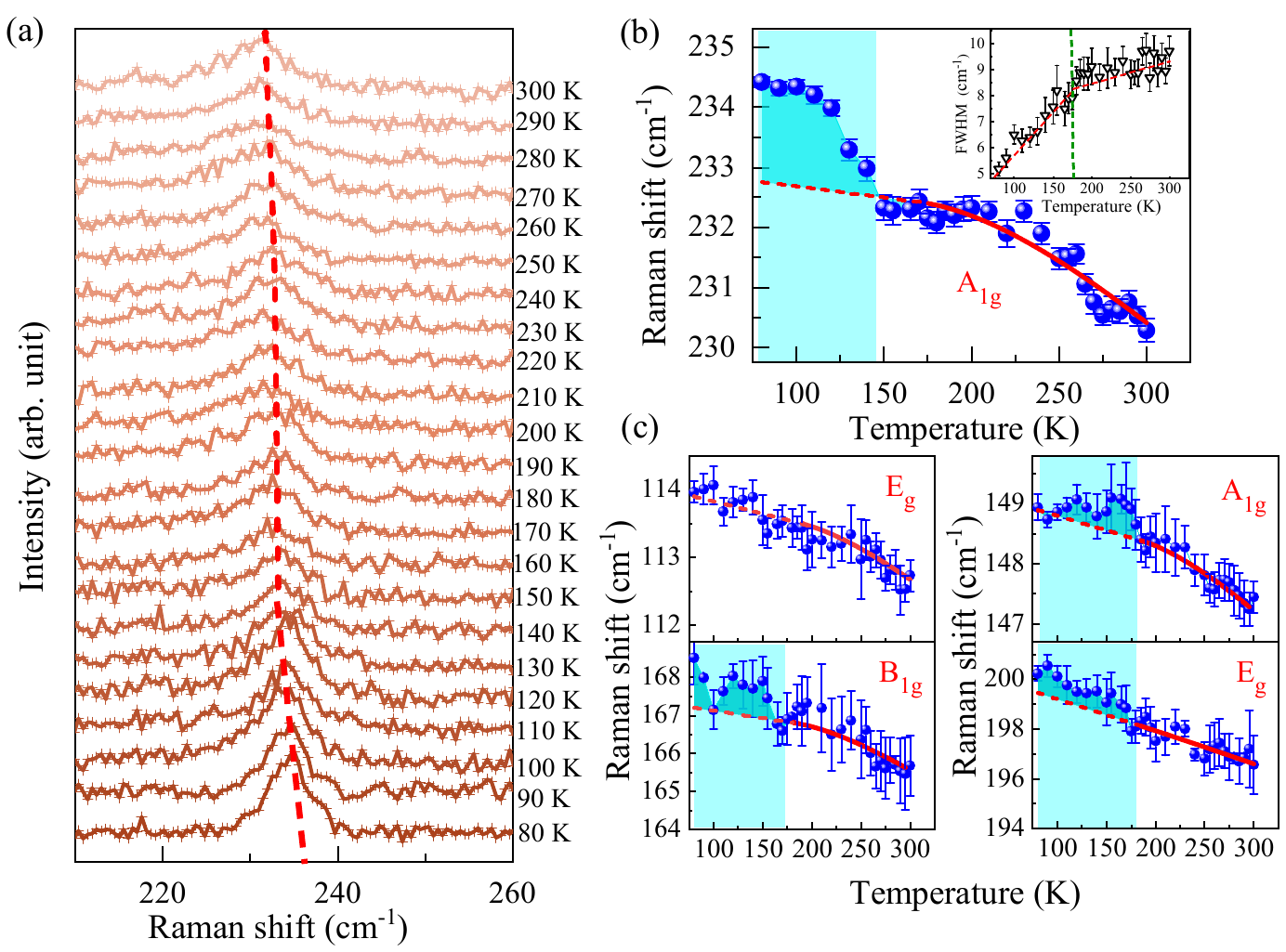}
\caption{(a) Characteristic temperature dependent Raman spectra of  EuCu$_{2}$As$_{2}$ within a temperature range between $80$~K and $300$~K. The red dashed line is a guide to the eyes (b) Temperature evolution of the Raman shift of the A$_{1g}$ mode (P7) of EuCu$_{2}$As$_{2}$. (Inset) Variation of FWHM of the same mode with temperature. The red dashed line in the  plot is a guide to the eye. The error bars are the standard deviation as obtained from the fitting procedure. (c) Temperature evolution of Raman shift of modes P1, P3, P4 and P5. The red solid curves in panels (b) and (c) are best fits to the data points using Eqn.~\ref{anh} between 160 and 300 K(see text). The fitted curves are extrapolated down to $80$~K and shown by dashed curves, with the temperature range being indicated by light cyan shaded areas. The region between the extrapolated curves and the experimental data points within the shaded area are filled with dark cyan. }
\label{fig.raman_evol}
\end{figure*}

\subsection{Phonon anomalies associated with the Raman modes}
\label{sec.temp}

To investigate the influence of the observed structural symmetry on the physical characteristics of the system, we discuss  the thermal evolution of the Raman spectra.
Fig.~\ref{fig.raman_evol}(a) presents the background subtracted magnified Raman spectra of EuCu$_2$As$_2$ focussing the A$_{1g}$ mode at $\sim$ 232 cm$^{-1}$, at various temperatures between $80$~K and $300$~K (full range Raman spectra are presented in Fig.~SF3 in the SM). 
It is to be noted that we do not observe any new spectral feature confirming the absence of any first-order structural phase transition, similar to other Cu-based 122 systems, over the temperature range of interest.

Each spectrum in Fig.~\ref{fig.raman_evol}(a) was deconvoluted with a Lorentzian function. The temperature evolution of Raman shift of the strongest A$_{1g}$ mode is presented in Fig.~\ref{fig.raman_evol}(b). Other modes were also analysed in a similar fashion. The thermal evolution of the next four most intense modes are shown in Fig.~\ref{fig.raman_evol}(c). The intensity of the modes P2 and P6 were too low to track their evolution reliably with temperature.

In solid, the evolution of Raman shift with temperature is mostly governed by change in anharmonic potential, if the change in volume of the unit cell is negligible with temperature and other contributions from spin, orbital, charge degrees of freedom do not couple with the phonon modes. 
Considering the four-phonon anharmonic decay process of phonon to two or more acoustic phonon modes, the change in Raman shift with temperature due to anharmonicity follows the relation~\cite{balkanski.wallis.83}: 
\begin{eqnarray}
\label{anh}
\omega(T) = \omega_o &+& A\left( 1+\frac{2}{e^{\hbar\omega_o/2K_BT}-1}\right) \\
\nonumber &+& B\left( 1+\frac{3}{e^{\hbar\omega_o/3K_BT}-1}\right) ,
\end{eqnarray}
where $\omega_0$ defines the Raman wavenumber at $T = 0$~K, and $A$ and $B$ are anharmonic constants. Reasonable fits to Eq.~(\ref{anh}) were obtained from $300$~K down to $165$~K for all the peaks shown in Fig.~\ref{fig.raman_evol}. 

As is clearly visible in Fig.~\ref{fig.raman_evol}(b), P7 deviates from the expected anharmonic behavior at $165$~K, with the mode frequency abruptly blueshifting around that temperature. 
Such a behavior is also observed for the other high intense modes, as shown in Fig.~\ref{fig.raman_evol}(c), although with lesser degrees (see the dark cyan shaded area within the light cyan band). 
Deviation in the thermal evolution of the peak shift, as described by Eq.~(\ref{anh}), can be attributed to the presence of some extraneous phase, electronic or magnetic in origin, which couples directly with the observed optical phonon modes affecting their decay channels. 
If the area between the experimentally measured data points and the extrapolated anharmonic fits between $80$~K and $165$~K can be taken to be a quantitative measure of the deviation, then it is interesting to observe that there is a monotonically decreasing trend from P7 to P1 (Fig.~SF4 in the SM). 
This indicates that the dominant instability responsible for this anomalous behavior primarily couples to the out of plane vibrations of the inter layer Cu(2) and As(1) atoms of the A$_{1g}$ mode.\\

\noindent{\it Possible explanation for the phonon anomaly:}
There can be a wide range of physical situations where the deviation from the standard anharmonic behavior described by Eq.~(\ref{anh}) can be observed. 
For example, coupling of phonons with magnons resulting from the cross-talks between multiple magnetic sublattices of correlated oxide systems would also lead to hardening of the Raman modes beyond the expected behaviour~\cite{Weber.Mad.22,Hill.Heather.19}. 
However, these are fundamentally insulating systems while EuCu$_2$As$_2$ is metallic with Eu$^{2+}$ contributing to only one localized magnetic lattice with no signs of frustration. 
Thus, it can be reasonably concluded that spin-phonon coupling of this form is at least not responsible for the present observed behavior.
Digression from anharmonicity in phonon frequency shift and linewidth also occurs due to interband scattering of Dirac states with optical phonons, which are found in many metallic binary arsenides like TaAs and Cd$_3$As$_2$~\cite{Sharafeev.Lemmens.17}. 
Although there are some qualitative similarities with what is observed here, there are fundamental differences between these systems. 
Firstly, in EuCu$_2$As$_2$, there are no evidences of linearly dispersing electronic states with topological properties~\cite{anand.johnston.15} and secondly, there are no signs of enhancement of specific heat capacity at low temperatures arising due to electronic quasielastic scattering~\cite{Sharafeev.Lemmens.17}. 
Hence, it seems likely that some other mechanism must be responsible for the observed anomaly in the current case.

Having discounted several possible scenarios which would result in such a stiffening of the phonon mode, it is important to investigate the probable origin of such an effect in this system. 
At this juncture, it might serve well to remind ourselves that EuCu$_2$As$_2$ lacks local center of symmetry (Fig.~\ref{fig.crys}), has quasi-two-dimensional (2D) metallic blocks in the form of CuAs layers with enhanced inter layer covalency. 
These prerequisites are well suited for a modified charge distribution within the system. In fact, such conditions are satisfied in transition metal dichalcogenides where charge density waves (CDW)~\cite{Lin.Dongjing.20} states are well documented~\cite{Joshi.Jaydeep.19}. 
Recent Raman studies on such dichalcogenide systems, as well as kagome metallic systems~\cite{Liu.Gan.22} hosting such CDW states, have shown that there is a similar anomalous hardening of the A$_{1g}$ phonon frequencies close to the CDW transition temperature which in our case would be around $165$~K. 
Examples of such a CDW state in a prototypical 122 type having locally non centrosymmetric structure appears in SrPt$_2$As$_2$~\cite{Kawasaki.Shinji.15} where spin relaxation rates measurements using NMR has been used to investigate the density wave order.

There are several avenues via which CDW states can be realised and Fermi surface nesting is one of them. 
Generally, {\it P4/nmm}, which we now have established EuCu$_2$As$_2$ to be in, would be expected to have a more quasi-2D Fermi surface topology compared to that for {\it I4/mmm} making it more amenable for such nesting conditions. 
Fig. ~\ref{fig.raman_evol}(b) inset shows the linewidth of the A$_{1g}$ mode decreases at a faster rate below $165$~K, which is also consistent with a putative CDW arising from the enhanced nesting scenario. 
This is because it indicates that there is a reduction in the overall electron-phonon coupling, which would be expected due to CDW induced partial gaping of the Fermi surface~\cite{Liu.Gan.22}. 
There are, however, several telltale signatures, like appearance of amplitude modes, two phonon modes, zone folded modes, which should be observed in the Raman spectra for a conclusive determination of the CDW state. Unfortunately, we have hitherto not been able to identify these modes in our measurements over the spectral range of our study.

Till now, the role of the magnetic Eu$^{2+}$ sublattice has not yet been considered. A Curie Weiss fit to the inverse susceptibility shows a very slight deviation below $165$~K, as shown in Fig.~\ref{fig.raman1s}(b). 
Similar behavior has been observed in related 122 systems~\cite{rotter.tegel.08b,krellner.caroca.08}, where it has been ascribed to the transition to a spin density wave (SDW) phase. 
EuCu$_{2}$As$_{2}$, similarly like EuFe$_{2}$As$_{2}$~\cite{xiao.su.09}, realized the AFM order with Eu magnetic moments along $a$ direction~\cite{anand.johnston.15}.
From theoretical calculations, we found the magnetic moment of Eu as $\sim 6.9$~$\mu_{B}$ (independently by the magnetic order), which is close to the nominal value of Eu atom, i.e. $7$~$\mu_{B}$.
We  also found a  small non-compensated magnetic moment (around $0.01$~$\mu_{B}$) on Cu(2) and As(1), i.e. atoms close to Eu.
Additionally, the energy of the ground state with different magnetic orders and Eu magnetic moments direction is comparable.
This suggests strong magnetic instability of the system.
Indeed, the AFM state can be easily altered to FM state, e.g., by application of  pressure~\cite{sangupta.alzamora.12}. Even at temperatures larger than the N'eel temperature ($\sim 15$~K), the magnetic fluctuations can affect the vibrational modes.
In the case mentioned for the A$_\text{1g}$ mode, during oscillations of the Cu(2) and As(1) atoms, the distance between them changes.
This can lead to modification of the interatomic force constant due to the magnetic interactions of the non-compensated magnetic moments of these atoms, in the SDW-like phase. 
In this context, it is worth mentioning that theoretically, it is possible to have an exchange instability wave where correlation among the itinerant electrons lead to a mixed CDW-SDW phase~\cite{Overhauser.68}. 
In such situations, the phase of the spin-up and spin-down density distributions would acquire an intermediate value between zero and $\pi/2$.

Thus, although it is probably not possible, at this juncture, to conclusively pinpoint the origin of the observed anomaly in the Raman mode frequency shift around $165$~K, there are reasonable evidences which indicate an underlying electronic wave instability to which there is commendable coupling of the optical phonons. 
More detailed studies are necessary to elucidate the nature of such a coupling.

\section{Summary}
\label{sec.summary}

In this paper, we present experimentally obtained Raman spectra and theoretical investigation of the lattice dynamics of EuCu$_{2}$As$_{2}$. 
The Raman spectrum of EuCu$_{2}$As$_{2}$ is much ``richer'' than that expected for a system with {\it I4/mmm} symmetry. 
From this, we expect that this compound does not possess the same crystal structure. 
Contrary to the initial study of EuCu$_{2}$As$_{2}$, which suggest realization of {\it I4/mmm} symmetry, like in BaCu$_{2}$As$_{2}$, we suggest realization of {\it P4/nmm} structure.
To support this statement, we perform theoretical analysis of the lattice dynamics of EuCu$_{2}$As$_{2}$ with {\it I4/mmm} and {\it P4/nmm} symmetries.
The phonon dispersion curve for {\it I4/mmm} possess imaginary soft modes, which show that this compound cannot be stable in this structure.
Contrary to this, the phonon spectrum for {\it P4/nmm} has only positive frequency modes (i.e., system is dynamically stable).
Additionally, from analysis of the modes at $\Gamma$ point, we found the Raman active modes, whose frequencies are in agreement with this observed experimentally.

Thermal evolution of the observed Raman modes exhibit an interesting deviation from the expected anharmonic behavior for optical phonons.
Below $165$~K there is significant hardening of the $230$~cm$^{-1}$ A$_{1g}$ mode, though similar behaviour has been observed for the other Raman modes as well. 
Taking into account the fact that EuCu$_{2}$As$_{2}$ is an $sp$ band metal with very little correlation, we see that there might be a subtle electronic density wave that couples effectively to the phonons. 
While it is not possible to conclusively establish the origin of such an anomaly, evidences suggest the presence of an electronic density wave instability. 
We notice that there are relatively very few reports on examples of CDW in the 122 phase~\cite{Kawasaki.Shinji.15}. 
Thus, the present work can serve to initiate more detailed studies to understand the role of local noncentrosymmetricity in the context of CDW and other related phases.

\section*{Acknowledgments}

Some figures were prepared using VESTA~\cite{21} software.
S.D.D. acknowledges Th. Doert and J. P. F. Jemetio in crystal growth. 
S.D.D. also acknowledges the financial support of ISIRD, IIT Kharagpur and SERB through grant No. ECR/2017/003083. 
A.R. and SDD acknowledge the financial support of SERB India through grant No. CRG/2021/000718. SDD is thankful to Professor E. V. Sampathkumaran for helpful discussions.
We kindly acknowledge support by National Science Centre (NCN, Poland) under Project No.~2021/43/B/ST3/02166 (A.P). G.V. acknowledges
Institute of Eminence University of Hyderabad (UoH-IoERC3-
21-046) for funding and CMSD University of Hyderabad
for providing the computational facility. MDS acknowledge PMRF for financial support.

\section*{References}

\bibliography{IOPEXPORT_BIB.bib}

\providecommand{\newblock}{}
\begin{thebibliography}{10}
\expandafter\ifx\csname url\endcsname\relax
  \def\url#1{{\tt #1}}\fi
\expandafter\ifx\csname urlprefix\endcsname\relax\def\urlprefix{URL }\fi
\providecommand{\eprint}[2][]{\url{#2}}

\bibitem{steglihc.aarts.79}
Steglich F, Aarts J, Bredl C~D, Lieke W, Meschede D, Franz W and Sch\"afer H 1979 {\em Phys. Rev. Lett.\/} {\bf 43} 1892 \urlprefix\url{https://doi.org/10.1103/PhysRevLett.43.1892}

\bibitem{Stockert2011}
Stockert O, Arndt J, Faulhaber E, Geibel C, Jeevan H~S, Kirchner S, Loewenhaupt M, Schmalzl K, Schmidt W, Si Q and Steglich F 2011 {\em Nat. Phys.\/} {\bf 7} 119 \urlprefix\url{https://doi.org/10.1038/nphys1852}

\bibitem{article}
Ni N, Nandi S, Kreyssig A, Goldman A, Mun E, Bud'ko S and Canfield P 2008 {\em Physical Review B\/} {\bf 78} 014523

\bibitem{Li_2020}
Li L, Parker D~S, Gai Z, Cao H~B and Sefat A~S 2020 {\em J. Phys.: Condens. Matter\/} {\bf 32} 295602 \urlprefix\url{https://doi.org/10.1088/1361-648X/ab7e60}

\bibitem{kudo.nishikubo.10}
Kudo K, Nishikubo Y and Nohara M 2010 {\em J. Phys. Soc. Jpn.\/} {\bf 79} 123710 \urlprefix\url{https://doi.org/10.1143/JPSJ.79.123710}

\bibitem{nagano.araoka.13}
Nagano Y, Araoka N, Mitsuda A, Yayama H, Wada H, Ichihara M, Isobe M and Ueda Y 2013 {\em J. Phys. Soc. Jpn.\/} {\bf 82} 064715 \urlprefix\url{https://doi.org/10.7566/JPSJ.82.064715}

\bibitem{mu.pan.18}
Mu Q, Pan B, Ruan B, Liu T, Zhao K, Shan L, Chen G and Ren Z 2018 {\em Sci. China Phys. Mech. Astron.\/} {\bf 61} 127409 \urlprefix\url{https://doi.org/10.1007/s11433-018-9285-5}

\bibitem{kim.landaeta.21}
Khim S, Landaeta J~F, Banda J, Bannor N, Brando M, Brydon P~M~R, Hafner D, Küchler R, Cardoso-Gil R, Stockert U, Mackenzie A~P, Agterberg D~F, Geibel C and Hassinger E 2021 {\em Science\/} {\bf 373} 1012 \urlprefix\url{https://doi.org/10.1126/science.abe7518}

\bibitem{rotter.tegel.08}
Rotter M, Tegel M and Johrendt D 2008 {\em Phys. Rev. Lett.\/} {\bf 101} 107006 \urlprefix\url{https://doi.org/10.1103/PhysRevLett.101.107006}

\bibitem{sasmal.lorenz.08}
Sasmal K, Lv B, Lorenz B, Guloy A~M, Chen F, Xue Y~Y and Chu C~W 2008 {\em Phys. Rev. Lett.\/} {\bf 101} 107007 \urlprefix\url{https://doi.org/10.1103/PhysRevLett.101.107007}

\bibitem{torikachvili.budko.08}
Torikachvili M~S, Bud'ko S~L, Ni N and Canfield P~C 2008 {\em Phys. Rev. Lett.\/} {\bf 101} 057006 \urlprefix\url{https://doi.org/10.1103/PhysRevLett.101.057006}

\bibitem{zocco.grube.13}
Zocco D~A, Grube K, Eilers F, Wolf T and L\"ohneysen H~v 2013 {\em Phys. Rev. Lett.\/} {\bf 111} 057007 \urlprefix\url{https://doi.org/10.1103/PhysRevLett.111.057007}

\bibitem{cho.yang.17}
Cho C~W, Yang J~H, Yuan N~F~Q, Shen J, Wolf T and Lortz R 2017 {\em Phys. Rev. Lett.\/} {\bf 119} 217002 \urlprefix\url{https://doi.org/10.1103/PhysRevLett.119.217002}

\bibitem{lin.miller.14}
Lin Q, Miller G~J and Corbett J~D 2014 {\em Inorg. Chem.\/} {\bf 53} 5875 \urlprefix\url{https://doi.org/10.1021/ic402991d}

\bibitem{wang.botti.21}
Wang H~C, Botti S and Marques M~A~L 2021 {\em npj Comput. Mater.\/} {\bf 7} 12 \urlprefix\url{https://doi.org/10.1038/s41524-020-00481-6}

\bibitem{singh.09}
Singh D~J 2009 {\em Phys. Rev. B\/} {\bf 79} 153102 \urlprefix\url{https://doi.org/10.1103/PhysRevB.79.153102}

\bibitem{wu.richard.15}
Wu S~F, Richard P, van Roekeghem A, Nie S~M, Miao H, Xu N, Qian T, Saparov B, Fang Z, Biermann S, Sefat A~S and Ding H 2015 {\em Phys. Rev. B\/} {\bf 91} 235109 \urlprefix\url{https://doi.org/10.1103/PhysRevB.91.235109}

\bibitem{Das.Craco.15}
Das S~D, Laad M~S, Craco L, Gillett J, Tripathi V and Sebastian S~E 2015 {\em Phys. Rev. B\/} {\bf 92} 155112 \urlprefix\url{https://doi.org/10.1103/PhysRevB.92.155112}

\bibitem{ptok.kapcia.21}
Ptok A, Kapcia K~J, Jochym P~T, \L{}a\.{z}ewski J, Ole\'{s} A~M and Piekarz P 2021 {\em Phys. Rev. B\/} {\bf 104} L041109 \urlprefix\url{https://doi.org/10.1103/PhysRevB.104.L041109}

\bibitem{Hafner.Khanenko.22}
Hafner D, Khanenko P, Eljaouhari E~O, K\"uchler R, Banda J, Bannor N, L\"uhmann T, Landaeta J~F, Mishra S, Sheikin I, Hassinger E, Khim S, Geibel C, Zwicknagl G and Brando M 2022 {\em Phys. Rev. X\/} {\bf 12} 011023 \urlprefix\url{https://doi.org/10.1103/PhysRevX.12.011023}

\bibitem{sengupta.paulose.05}
Sengupta K, Paulose P~L, Sampathkumaran E~V, Doert T and Jemetio J~P~F 2005 {\em Phys. Rev. B\/} {\bf 72} 184424 \urlprefix\url{https://doi.org/10.1103/PhysRevB.72.184424}

\bibitem{sangupta.alzamora.12}
Sengupta K, Alzamora M, Fontes M~B, Sampathkumaran E~V, Ramos S~M, Hering E~N, Saitovitch E~M~B, Paulose P~L, Ranganathan R, Doert T and Jemetio J~P~F 2012 {\em J. Phys.: Condens. Matter\/} {\bf 24} 096004 \urlprefix\url{https://doi.org/10.1088/0953-8984/24/9/096004}

\bibitem{anand.johnston.15}
Anand V~K and Johnston D~C 2015 {\em Phys. Rev. B\/} {\bf 91} 184403 \urlprefix\url{https://doi.org/10.1103/PhysRevB.91.184403}

\bibitem{miclea.nicklas.09}
Miclea C~F, Nicklas M, Jeevan H~S, Kasinathan D, Hossain Z, Rosner H, Gegenwart P, Geibel C and Steglich F 2009 {\em Phys. Rev. B\/} {\bf 79} 212509 \urlprefix\url{https://ldoi.org/10.1103/PhysRevB.79.212509}

\bibitem{Adhikary_2013}
Adhikary G, Sahadev N, Biswas D, Bindu R, Kumar N, Thamizhavel A, Dhar S~K and Maiti K 2013 {\em J. Phys.: Condens. Matter\/} {\bf 25} 225701 \urlprefix\url{https://doi.org/10.1088/0953-8984/25/22/225701}

\bibitem{blochl.94}
Bl\"ochl P~E 1994 {\em Phys. Rev. B\/} {\bf 50} 17953 \urlprefix\url{https://doi.org/10.1103/PhysRevB.50.17953}

\bibitem{kresse.hafner.94}
Kresse G and Hafner J 1994 {\em Phys. Rev. B\/} {\bf 49} 14251 \urlprefix\url{https://doi.org/10.1103/PhysRevB.49.14251}

\bibitem{perdew.ruzsinszky.08}
Perdew J~P, Ruzsinszky A, Csonka G~I, Vydrov O~A, Scuseria G~E, Constantin L~A, Zhou X and Burke K 2008 {\em Phys. Rev. Lett.\/} {\bf 100} 136406 \urlprefix\url{https://doi.org/10.1103/PhysRevLett.100.136406}

\bibitem{dudarev.botton.98}
Dudarev S~L, Botton G~A, Savrasov S~Y, Humphreys C~J and Sutton A~P 1998 {\em Phys. Rev. B\/} {\bf 57} 1505 \urlprefix\url{https://doi.org/10.1103/PhysRevB.57.1505}

\bibitem{monkhorst.pack.76}
Monkhorst H~J and Pack J~D 1976 {\em Phys. Rev. B\/} {\bf 13} 5188 \urlprefix\url{https://doi.org/10.1103/PhysRevB.13.5188}

\bibitem{stokes.hatch.05}
Stokes H~T and Hatch D~M 2005 {\em J. Appl. Cryst.\/} {\bf 38} 237 \urlprefix\url{https://doi.org/10.1107/S0021889804031528}

\bibitem{togo.tanaka.18}
Togo A and Tanaka I 2018 {\sc Spglib}: a software library for crystal symmetry search (\textit{Preprint} \eprint{arXiv:1808.01590})

\bibitem{hinuma.pizzi.17}
Hinuma Y, Pizzi G, Kumagai Y, Oba F and Tanaka I 2017 {\em Comput. Mater. Sci.\/} {\bf 128} 140 \urlprefix\url{https://doi.org/10.1016/j.commatsci.2016.10.015}

\bibitem{parlinski.li.97}
Parlinski K, Li Z~Q and Kawazoe Y 1997 {\em Phys. Rev. Lett.\/} {\bf 78} 4063 \urlprefix\url{https://doi.org/10.1103/PhysRevLett.78.4063}

\bibitem{tadano.gohda.14}
Tadano T, Gohda Y and Tsuneyuki S 2014 {\em J. Phys.: Condens. Matter\/} {\bf 26} 225402 \urlprefix\url{https://doi.org/10.1088/0953-8984/26/22/225402}

\bibitem{togo.chaput.23}
Togo A, Chaput L, Tadano T and Tanaka I 2023 {\em J. Phys. Condens. Matter\/} {\bf 35} 353001

\bibitem{dunner.mewis.95}
D\"{u}nner J, Mewis A, Roepke M and Michels G 1995 {\em Z. anorg. allg. Chem.\/} {\bf 621} 1523 \urlprefix\url{https://doi.org/10.1002/zaac.19956210915}

\bibitem{das.mcfadden.10}
Das S, McFadden K, Singh Y, Nath R, Ellern A and Johnston D~C 2010 {\em Phys. Rev. B\/} {\bf 81} 054425 \urlprefix\url{https://ldoi.org/10.1103/PhysRevB.81.054425}

\bibitem{rotter.tegel.08b}
Rotter M, Tegel M, Johrendt D, Schellenberg I, Hermes W and P\"ottgen R 2008 {\em Phys. Rev. B\/} {\bf 78} 020503(R) \urlprefix\url{https://doi.org/10.1103/PhysRevB.78.020503}

\bibitem{krellner.caroca.08}
Krellner C, Caroca-Canales N, Jesche A, Rosner H, Ormeci A and Geibel C 2008 {\em Phys. Rev. B\/} {\bf 78} 100504(R) \urlprefix\url{https://doi.org/10.1103/PhysRevB.78.100504}

\bibitem{litvinchuk.hadjiev.08}
Litvinchuk A~P, Hadjiev V~G, Iliev M~N, Lv B, Guloy A~M and Chu C~W 2008 {\em Phys. Rev. B\/} {\bf 78} 060503(R) \urlprefix\url{https://doi.org/10.1103/PhysRevB.78.060503}

\bibitem{wdowik.jaglo.15}
Wdowik U~D, Jag\l{}o G and Piekarz P 2015 {\em J. Phys.: Condens. Matter\/} {\bf 27} 415403 \urlprefix\url{https://doi.org/10.1088/0953-8984/27/41/415403}

\bibitem{baum.li.18}
Baum A, Li Y, Tomi\'{c} M, Lazarevi\'{c} N, Jost D, L\"offler F, Muschler B, B\"ohm T, Chu J~H, Fisher I~R, Valent\'{\i} R, Mazin I~I and Hackl R 2018 {\em Phys. Rev. B\/} {\bf 98} 075113 \urlprefix\url{https://doi.org/10.1103/PhysRevB.98.075113}

\bibitem{ptok.sternik.19}
Ptok A, Sternik M, Kapcia K~J and Piekarz P 2019 {\em Phys. Rev. B\/} {\bf 99} 134103 \urlprefix\url{https://doi.org/10.1103/PhysRevB.99.134103}

\bibitem{balkanski.wallis.83}
Balkanski M, Wallis R~F and Haro E 1983 {\em Phys. Rev. B\/} {\bf 28} 1928 \urlprefix\url{https://doi.org/10.1103/PhysRevB.28.1928}

\bibitem{Weber.Mad.22}
Weber M~C, Guennou M, Evans D~M, Toulouse C, Simonov A, Kholina Y, Ma X, Ren W, Cao S, Carpenter M~A, Dkhil B, Fiebig M and Kreisel J 2022 {\em Nature Commun.\/} {\bf 13} 443 \urlprefix\url{https://doi.org/10.1038/s41467-021-27267-8}

\bibitem{Hill.Heather.19}
Hill H~M, Chowdhury S, Simpson J~R, Rigosi A~F, Newell D~B, Berger H, Tavazza F and Hight~Walker A~R 2019 {\em Phys. Rev. B\/} {\bf 99} 174110 \urlprefix\url{https://doi.org/10.1103/PhysRevB.99.174110}

\bibitem{Sharafeev.Lemmens.17}
Sharafeev A, Gnezdilov V, Sankar R, Chou F~C and Lemmens P 2017 {\em Phys. Rev. B\/} {\bf 95} 235148 \urlprefix\url{https://doi.org/10.1103/PhysRevB.95.235148}

\bibitem{Lin.Dongjing.20}
Lin D, Li S, Wen J, Berger H, Forr{\'o} L, Zhou H, Jia S, Taniguchi T, Watanabe K, Xi X and Bahramy M~S 2020 {\em Nat. Commun.\/} {\bf 11} 2406 \urlprefix\url{https://doi.org/10.1038/s41467-020-15715-w}

\bibitem{Joshi.Jaydeep.19}
Joshi J, Hill H~M, Chowdhury S, Malliakas C~D, Tavazza F, Chatterjee U, Hight~Walker A~R and Vora P~M 2019 {\em Phys. Rev. B\/} {\bf 99} 245144 \urlprefix\url{https://doi.org/10.1103/PhysRevB.99.245144}

\bibitem{Liu.Gan.22}
Liu G, Ma X, He K, Li Q, Tan H, Liu Y, Xu J, Tang W, Watanabe K, Taniguchi T, Gao L, Dai Y, Wen H~H, Yan B and Xi X 2022 {\em Nature Commun.\/} {\bf 13} 3461 \urlprefix\url{https://doi.org/10.1038/s41467-022-31162-1}

\bibitem{Kawasaki.Shinji.15}
Kawasaki S, Tani Y, Mabuchi T, Kudo K, Nishikubo Y, Mitsuoka D, Nohara M and Zheng G~q 2015 {\em Phys. Rev. B\/} {\bf 91} 060510(R) \urlprefix\url{https://doi.org/10.1103/PhysRevB.91.060510}

\bibitem{xiao.su.09}
Xiao Y, Su Y, Meven M, Mittal R, Kumar C~M~N, Chatterji T, Price S, Persson J, Kumar N, Dhar S~K, Thamizhavel A and Brueckel T 2009 {\em Phys. Rev. B\/} {\bf 80} 174424 \urlprefix\url{https://doi.org/10.1103/PhysRevB.80.174424}

\bibitem{Overhauser.68}
Overhauser A~W 1968 {\em Phys. Rev.\/} {\bf 167} 691 \urlprefix\url{https://doi.org/10.1103/PhysRev.167.691}

\bibitem{21}
Momma K and Izumi F 2011 {\em J. Appl. Crystallogr.\/} {\bf 44} 1272 \urlprefix\url{http://doi.org/10.1107/S0021889811038970}

\end{thebibliography}

\end{document}



\begin{center}
\textbf{\Large Supplemental Material}\\[.3cm]
\textbf{\large Evidences for local non-centrosymmetricity \\and strong phonon anomaly in EuCu$_{2}$As$_{2}$:\\A Raman spectroscopy and lattice dynamics study}\\[.3cm]
D. Swain {\it et al.}\\[.2cm]
(Dated: \today)
\\[1cm]
\end{center}

\setcounter{equation}{0}
\renewcommand{\theequation}{SE\arabic{equation}}
\setcounter{figure}{0}
\renewcommand{\thefigure}{SF\arabic{figure}}
\setcounter{section}{0}
\renewcommand{\thesection}{SS\arabic{section}}
\setcounter{table}{0}
\renewcommand{\thetable}{ST\arabic{table}}
\setcounter{page}{1}


In this Supplemental Material, we present additional results:
\begin{itemize}
\item Sec.~\ref{sec.sm.samplesynthesis} -Sample growth details and basic characterisation
\item Sec.~\ref{sec.sm.IRactive} - atoms displacement by IR active modes
\item Sec.~\ref{sec.sm.temp_raman} - the temperature dependence of Raman spectra
\end{itemize}

\section{Sample growth details and basic characterisation}
 \label{sec.sm.samplesynthesis} 
 A polycrystalline sample of EuCu$_2$As$_2$ was synthesized by directly reacting the high pure elemental components, Eu powder (99.999\%), Cu powder (99.9\%), and As sponge (99.9\%) in a 1:2:2 molar ratio. The mixture was sealed in an evacuated silica tube, whose handling and sealing was done under Ar filled glove box to avoid oxidation. Then the ampoule was undergone heat treatment in  a box furnace. The heating process involved ramping the temperature up to 500 °C at a rate of 8 °C/min, followed by annealing at 900 °C for 10–12 days and cooled to room temperature. A polycrystalline pellet was obtained at the end of the process, which was stable in air~( see Ref. [21] of the manuscript). \\

\begin{figure}[h!]
    \centering
    \includegraphics[scale=0.65]{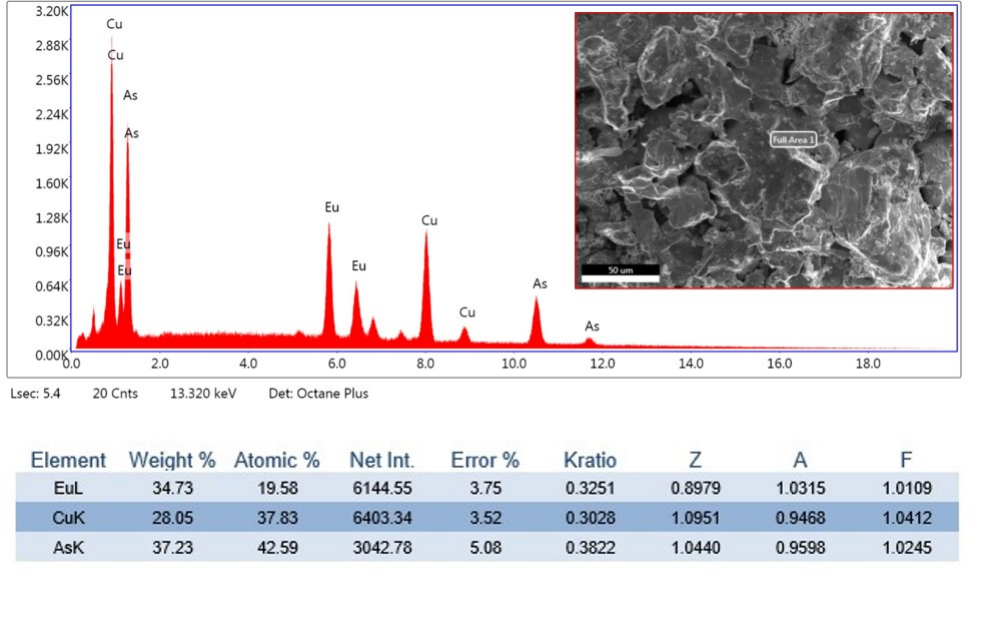}
    \caption{EDX spectra of EuCu$_2$As$_2$, (inset) SEM image of the focused sample. The table below shows the atomic percentage of the constituent elements}
    \label{Figure_SEM}
\end{figure}

Figure~\ref{Figure_SEM} gives the EDX spectrum of EuCu$_2$As$_2$ with the SEM image of the studied sample is given in the inset. The table given below the image shows the atomic percentage of constituent elements and it is consistent with the expected stoichiometric ratio of 1:2:2 (Eu:Cu:As).

\section{Atomic displacements by IR active modes}
\label{sec.sm.IRactive}

\begin{figure}[!h]
	\centering
	\includegraphics[scale=0.85]{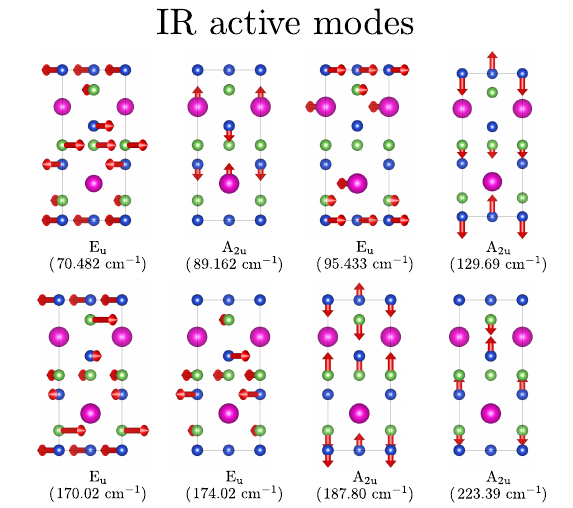}
	\caption{Atomic displacements realized by the IR active modes of  EuCu$_2$As$_2$ with {\it P4/nmm} symmetry.}
	\label{fig.raman4s}
\end{figure}

Figure~\ref{fig.raman4s} presents the atomic displacements by the IR active modes of EuCu$_2$As$_2$ with {\it P4/nmm} symmetry, as referred in the main text.

\section{Temperature dependent Raman spectrum}
\label{sec.sm.temp_raman} 
\begin{figure}[!h]
\centering
\includegraphics[scale=0.2]{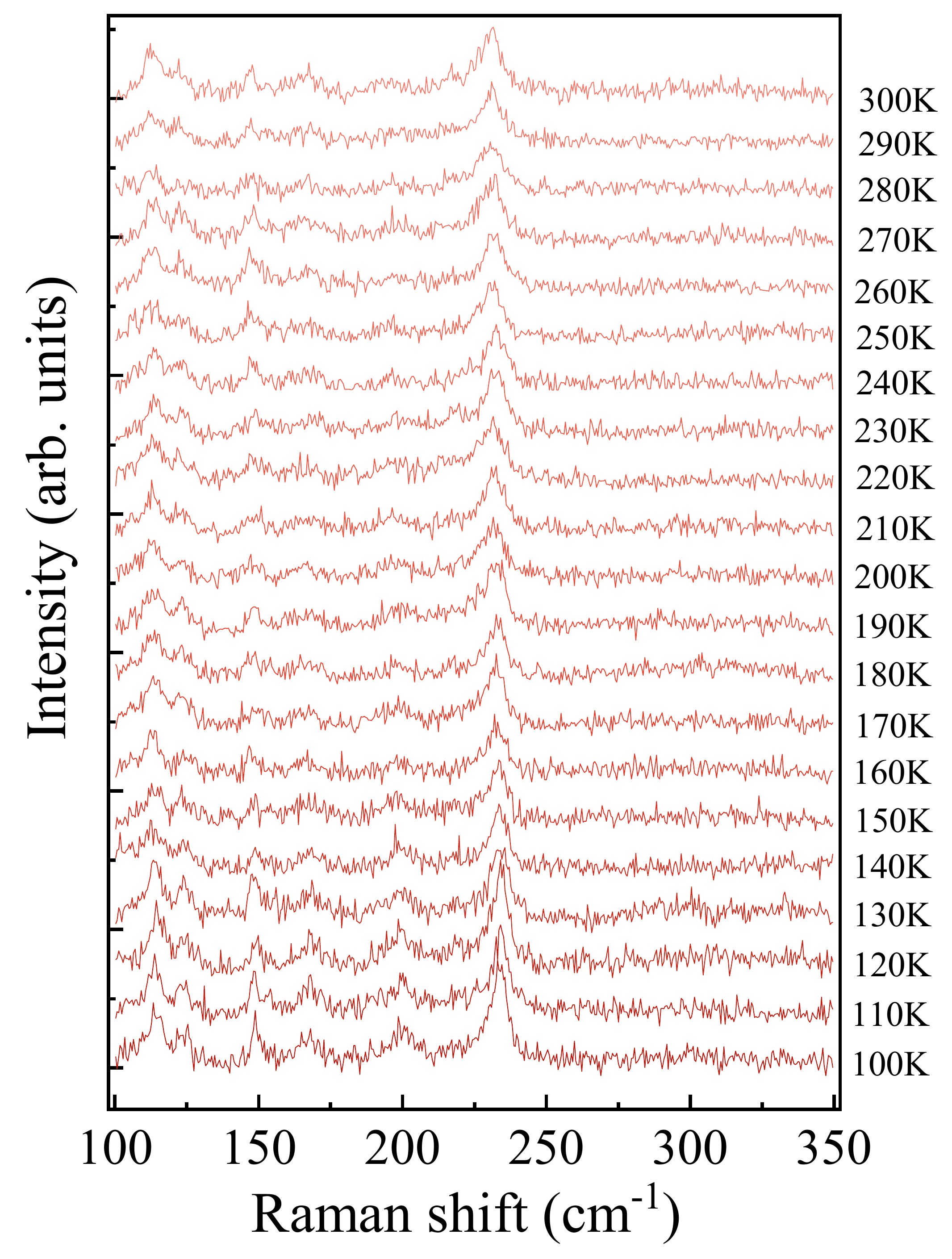}
\caption{ Temperature evolution of characteristics Raman spectra of EuCu$_2$As$_2$ between $100$~K and $300$~K.}
 \label{fig.raman2s}
\end{figure}

Fig.~\ref{fig.raman2s} represents the characteristic background subtracted Raman spectra of EuCu$_2$As$_2$ taken using wavelength of $532$~nm. 
These spectra are plotted within a spectral window of $100$--$350$~cm$^{-1}$ over the temperature range between $100$~K and $300$~K. 

\begin{figure}[h!]
\centering
\includegraphics[scale=0.4]{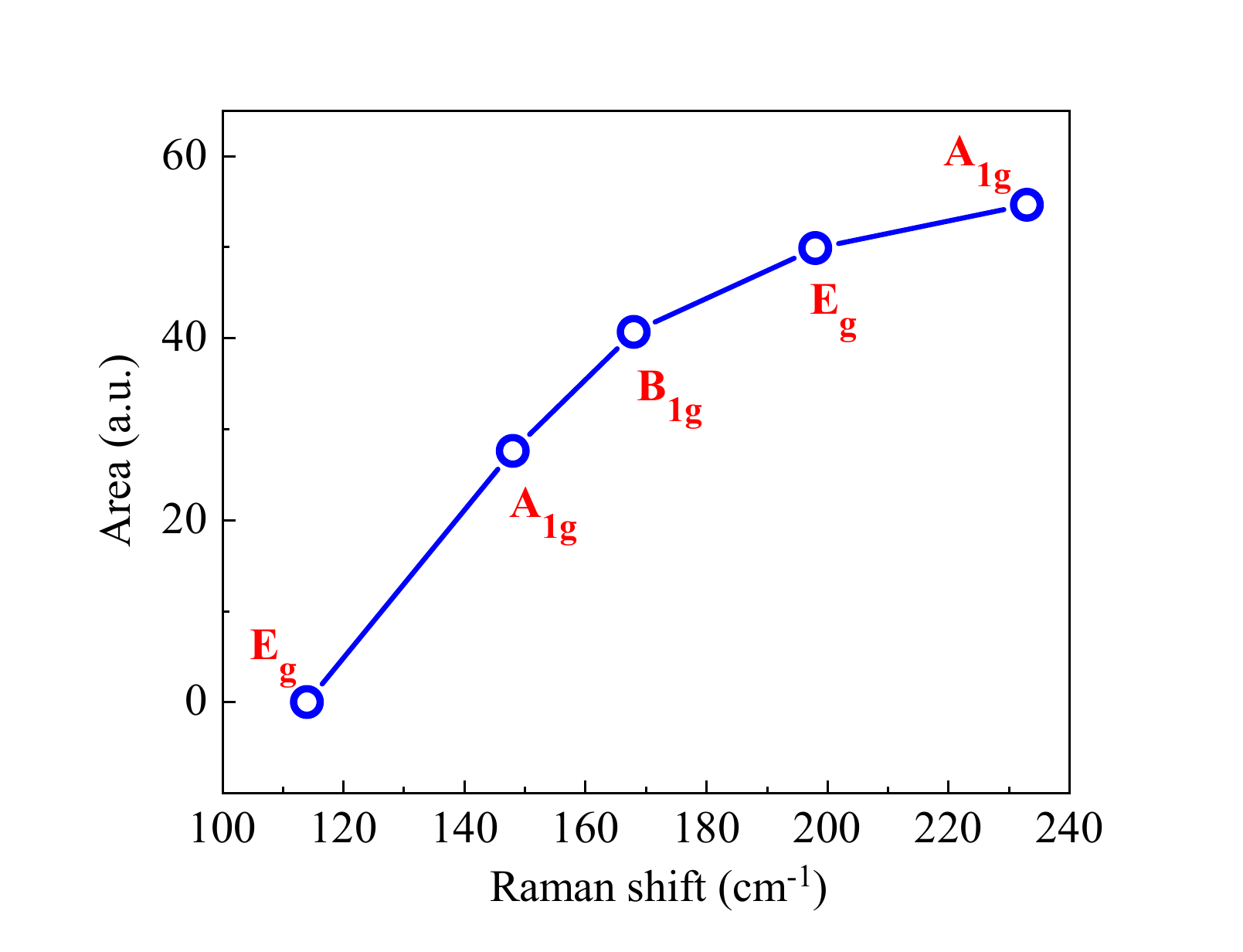}
\caption{Quantitative representation of the area obtained from the deviation of the obtained peak positions from anharmonicity curve fit.}
 \label{fig.raman3s}
\end{figure}

If we consider the area between the experimentally obtained data points and the extrapolated anharmonic fits within the temperature range of $80$~K to $165$~K, we can interpret this as a quantitative indicator of the deviation. Notably, there is an intriguing observation of a consistent decreasing trend from P7 to P1 (see Fig.~\ref{fig.raman3s}). \\